\documentclass[10pt,conference]{IEEEtran}
\IEEEoverridecommandlockouts
\usepackage{cite}
\usepackage{amsmath,amssymb,amsfonts}
\usepackage{algorithmic}
\usepackage{graphicx}
\usepackage{array}
\newcolumntype{P}[1]{>{\centering\arraybackslash}p{#1}}
\newcolumntype{M}[1]{>{\centering\arraybackslash}m{#1}}
\newcolumntype{L}[1]{>{\raggedright\arraybackslash}m{#1}}
\usepackage{textcomp}
\usepackage{xcolor}
\usepackage{float}
\def\BibTeX{{\rm B\kern-.05em{\sc i\kern-.025em b}\kern-.08em
    T\kern-.1667em\lower.7ex\hbox{E}\kern-.125emX}}
    
\begin{document}

\title{Automatic Live Music Song Identification Using Multi-level Deep Sequence Similarity Learning \\
}

\author{\IEEEauthorblockN{1\textsuperscript{st} Aapo Hakala}
\IEEEauthorblockA{\textit{Audio Research Group} \\
\textit{Tampere University}\\
Tampere, Finland \\
aapo.hakala@tuni.fi}

\and
\IEEEauthorblockN{2\textsuperscript{nd} Trevor Kincy}
\IEEEauthorblockA{\textit{Future Lab} \\
\textit{Utopia Music}\\
Zug, Switzerland \\
trevor.kincy@utopiamusic.com}

\and
\IEEEauthorblockN{3\textsuperscript{rd} Tuomas Virtanen}
\IEEEauthorblockA{\textit{Audio Research Group} \\
\textit{Tampere University}\\
Tampere, Finland \\
tuomas.virtanen@tuni.fi}
}

\maketitle

\begin{abstract}
This paper studies the novel problem of automatic live music song identification, where the goal is, given a live recording of a song, to retrieve the corresponding studio version of the song from a music database. We propose a system based on similarity learning and a Siamese convolutional neural network-based model. The model uses cross-similarity matrices of multi-level deep sequences to measure musical similarity between different audio tracks. A manually collected custom live music dataset is used to test the performance of the system with live music. The results of the experiments show that the system is able to identify 87.4\% of the given live music queries.
\end{abstract}

\begin{IEEEkeywords}
live song identification, music information retrieval, similarity learning, Siamese network, cross-similarity matrice, multi-level deep sequences
\end{IEEEkeywords}

\section{Introduction}\label{sec:introduction}
Humans can easily recognize a live version of a familiar song even if there are drastic musical and structural differences. Two versions of the same song can have differences in their instrumentation, arrangement, tempo, lyrics, song key, individual chords or entire chord progressions and even changes in song-defining melodies. Live music environment brings additional complexity to the task as crowd noises, crowd interactions, and instrumental improvisation will often occur. Automated live performance tracking systems that could deal with all the above-mentioned musical properties are still yet to be established. Such a system would greatly benefit music industry from songwriters to publishers and performance rights societies who must currently make reports of live performances manually in order to collect their performance rights royalties.

Automatic song identification systems have been developed and used extensively ever since the widespread use of smart phones. Acoustic fingerprinting has been used in applications like Youtube Content-ID and Shazam for preventing plagiarism and to provide information about songs played in the background \cite{content_id, saz}. Both systems have been proven to perform well in audio copy detection while also being robust to background noises. However, acoustic fingerprinting is not robust against considerable musical changes, which is why it is not applicable in live music use. The task of cover song identification (CSI) on the other hand is closer to the problem at hand and it is also a topic that has been studied more recently. The state-of-the-art methods in CSI are nowadays based on deep learning techniques, namely convolutional neural networks (CNNs) \cite{refe5, refe2, refe3, refe4}. It has been shown that deep neural network-based solutions outperform conventional methods that rely on various combinations of feature extraction and distance metrics \cite{refe_old1, refe_old2, refe_old3}. Despite the recent progress in CSI, there seems to be a gap in the literature when it comes to song identification in live music context. None of the recent studies provide information about their performance on live music data, although there has been more focus on the topic in the past with conventional methods \cite{refe_live1, refe_live2, refe_live3}. Given the superiority of deep learning techniques in CSI and the lack of research from live music perspective, we want to bring more attention on the subject by assessing the problem with our work.

\begin{figure*}[t]
 \centerline{
 \includegraphics[width=\textwidth]{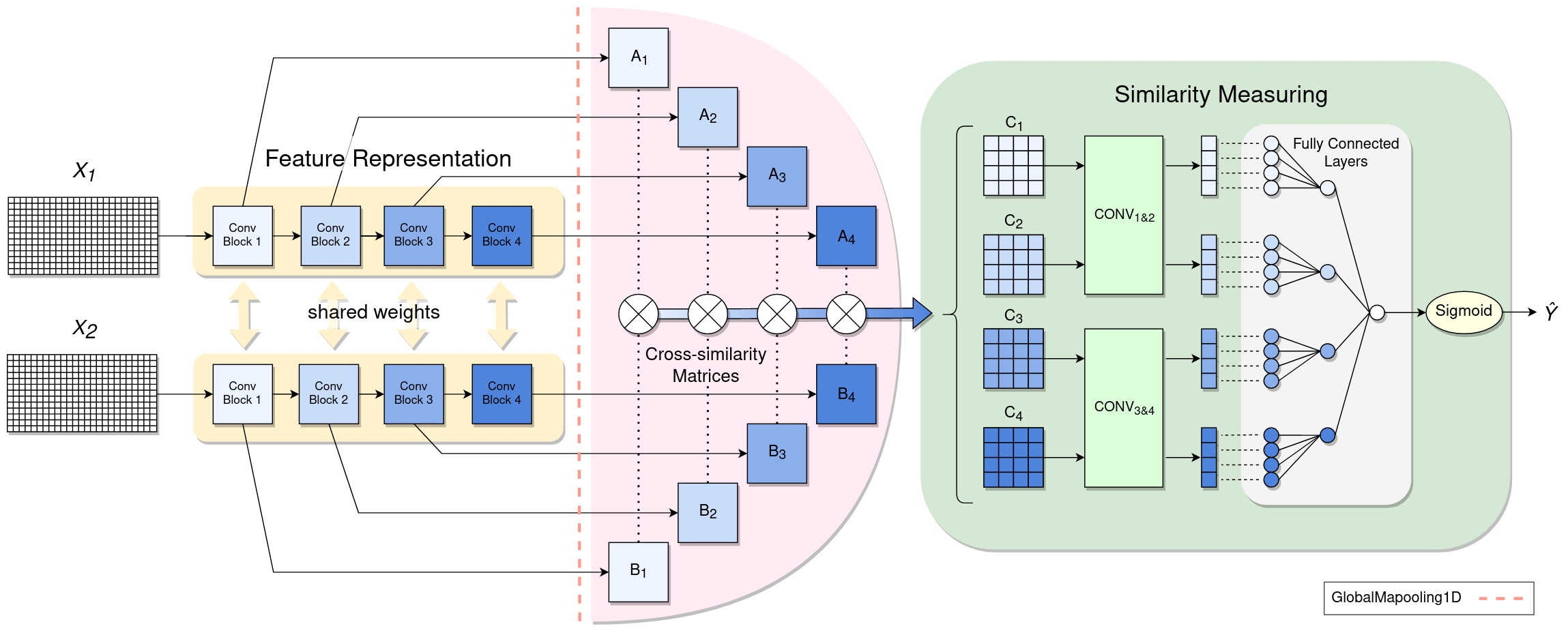}}
 \caption{A block diagram of the model architecture. In model input CQ-spectrograms $X_1$ and $X_2$ from two separate tracks are fed to different branches of a SCNN performing the feature representation. Resulting multi-level deep sequences $A_1 \dots A_4$ and $B_1 \dots B_4$ are then used to compute level-specific CSMs $C_1 \dots C_4$. Two parallel CNNs are used for similarity measuring, followed by four fully connected layers, a single output neuron and a sigmoid activation function. The final output is a similarity score $\hat{Y} \in [0, 1]$ predicting the similarity of the given input tracks as a probability value.}
 \label{fig:block_diag}
\end{figure*}

In this paper we study the problem of live music song identification. We present a system that applies similarity learning by training a siamese convolutional neural network (SCNN)-based model, which uses cross-similarity matrices (CSMs) of multi-level deep sequences as proposed originally by Jiang et al. \cite{refe1}. Three alternative feature extraction methods are designed and used for model training that are 1) basic method, 2) chorus alignment method and 3) crowd noise method. We collect a custom live music dataset to evaluate the performance of the developed system with live music data.

The structure of the paper is the following. Section \ref{sec:system} provides a detailed system description about the proposed method and the model architecture. Information about the data and the feature variations used in the experiments are then covered in section \ref{sec:data}. The results of the experiments and the observations about the system performance are presented in section \ref{sec:evaluation} followed by the final conclusions in section \ref{sec:conclusion}.

\section{Live music song identification system}\label{sec:system}


The goal of the system is to recognize songs from live recording tracks and retrieve their corresponding studio versions from a music database. The system comprises a music database and a model that measures musical and structural similarities between a given pair of audio tracks. To identify the song from a live recording, the system first uses the model to compute similarity scores between the unknown live recording track and each song in the database. The similarity scores are then used to form a relevance-based ranking of the database that determines which studio version has the highest similarity score and is finally retrieved as the predicted output.

The proposed model architecture is shown in Fig. \ref{fig:block_diag}. The model is a cascade of three jointly trained CNNs that makes use of cross-similarity matrices of multi-level deep sequences. The main benefits of this method are that no handcrafted features are needed, meaning that no information is lost during the feature extraction stage \cite{refe1}. The method is also robust against changes in tempo and key transposition, which is a crucial property when dealing with live music data.

The learning setup is done according to similarity learning by training the model with pairs of similar and non-similar objects. Here similar objects mean pairs of audio tracks containing the same underlying song with alternative presentations, while non-similar objects are pairs of tracks presenting two different musical compositions. Similar objects are referred as class 1 while non-similar objects are referred as class 0. Let training dataset $D$ consist of pairs of features of songs $(X_1, X_2)$, that are each given a ground truth label $Y \in \{1, 0\}$ to signify whether $X_1$ and $X_2$ present the same song or not, respectively. The goal in the training is then to minimize a \textit{Binary Cross Entropy} (BCE) loss. Let $\theta$, $\gamma_1$ and $\gamma_2$ denote the parametrization of the SCNN and the parallel CNNs in the similarity measuring accordingly. The objective in the training can now be written as
\begin{equation}
\arg \min_{\theta, \gamma_1, \gamma_2} \sum_{n}{-Y_n\log(\hat{Y_n}) - (1-Y_n)\log(1-\hat{Y_n})}\label{eq_loss},
\end{equation} 
where $\hat{Y} \in [0, 1]$ is a predicted similarity score produced as the model output and $n$ denotes the index of the paired song examples in the dataset $D$.

\subsection{Features}
\label{feat}
In feature extraction, a fixed duration of 120 seconds of audio is extracted from the beginning of each song. Songs with a duration less than two minutes are zero padded to have the same length. The audio tracks are downsampled to have a sampling rate of 22050 Hz and converted from stereo to mono signals by averaging samples across the channels. Constant-Q (CQ) spectrograms are then computed from the time-domain audio signals by taking the absolute values of CQ transforms. The frequency resolution of a CQ-spectrogram enables the distinction of adjacent notes in a chromatic scale even at high frequencies, thus making it a suitable feature presentation method for detecting melodic patterns and chord progressions \cite{constantQ}. For each CQ-spectrogram $X \in \mathbb{R} ^{N \times T}$ the number of frequency bins $N=72$ and the number of time-frames $T=401$. In CQ transform the number of frequency bins per octave is 12, the minimum frequency is 32.7 Hz and the hop length is 299 ms. The CQ-features of each song track are standardized separately to have a mean of 0 and a standard deviation of 1.

\begin{figure*}[h]
 \centerline{
 \includegraphics[width=0.95\textwidth]{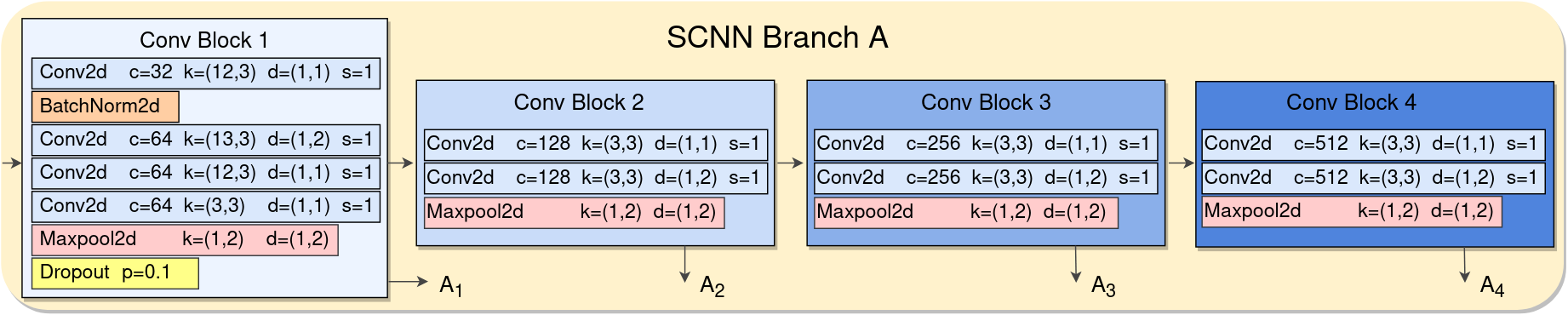}}
 \caption{The inner structure of branch $A$ of the SCNN. The data flow and the parametrization of the four convolution blocks are identical in branch $B$. The layer parameter names are abbreviated as follows: c=number of channels, k=kernel size, d=dilation, s=stride and p=dropout probability.}
 \label{fig:scnn}
\end{figure*}

\subsection{Feature Representation}
The SCNN in the model input is made of two identical CNN-branches $A$ and $B$ that share the same weights and parameters. At first, the input features $X_1$ and $X_2$ are fed to separate branches of the SCNN. The CNN-branches consist of four blocks of convolution and maxpooling operations as shown in Fig. \ref{fig:scnn}. Each convolution block produces a multi-level deep sequence $A_k = \{a_1^k, a_2^k, \dots, T_k^k\}$ or $B_k = \{b_1^k, b_2^k, \dots, T_k^k\}$, where $k \in \{1, 2, 3, 4\}$ denotes the level of deep sequence, $T_k$ is the width of a convolved CQ-spectrogram at $k$ and $A_k, B_k \in \mathbb{R}^{N_k \times T_k}$, where $N_k$ is the number of convolution channels at $k$. Initially the convolution blocks produce three-dimensional tensors having a shape of $N_k \times H_k \times T_k$, where $H_k$ is the height of the convolved CQ-spectrograms. At the end of the feature representation, a Global Maxpooling1d-operation is applied to each block output along the frequency axis, thus collapsing $H_k$ to $1$ and reducing the dimensions of $A_k$ and $B_k$ to $N_k \times T_k$. The use of deep sequences enables examining and comparing the input audio tracks at different abstraction levels. Multiple levels of granularity presented by the deep sequences provide more information for measuring the similarity between $X_1$ and $X_2$. More details about the inner structure of the SCNN are provided in Fig. \ref{fig:scnn}.

\subsection{Cross-similarity Matrices}
\begin{figure}[h]
 \centerline{
 \includegraphics[width=\columnwidth]{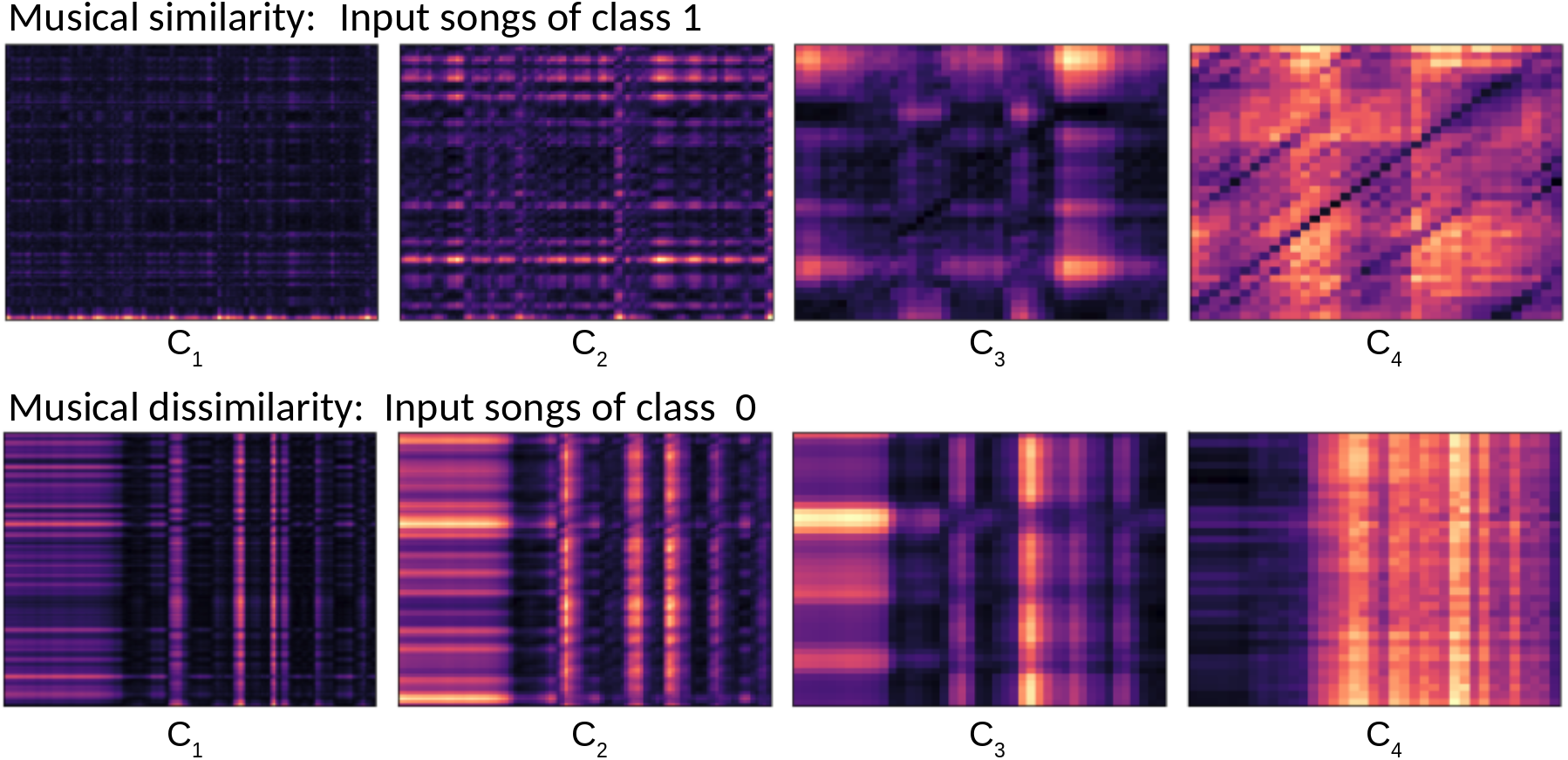}}
 \caption{Visualization of CSMs. The input songs used in class 1 example are 'Losing My Religion' by R.E.M and a live recording of the same song. In class 0 example the same live recording track is compared against 'E5150' by Black Sabbath.}
 \label{fig:csm}
\end{figure}

After extracting the multi-level deep sequences $A_k$ and $B_k$ the parallel branches of the SCNN are connected by computing CSMs between the convolved inputs. Level-specific CSMs $C_k \in \mathbb{R}^{T_k \times T_k}$ are calculated for each $k$ as
\begin{equation}
C_{k,(i,j)} = \|a_i^k - b_j^k\|^2,\label{eq_csm}
\end{equation}where $i, j \in \{  1, \dots, T_k\}$. The resulting CSMs can be viewed as rectangular images consisting of $T_k^2$ pixel values. Examples of such images are shown in Fig. \ref{fig:csm}. Any two songs with similar musical content will produce diagonal patterns to the CSMs as demonstrated by the first row of images in Fig. \ref{fig:csm}. Conversely, songs with dissimilar musical content result in grid-like shapes as seen on the bottom row of images.

\subsection{Similarity Measuring}
\begin{figure}[h]
 \centerline{
 \includegraphics[width=\columnwidth]{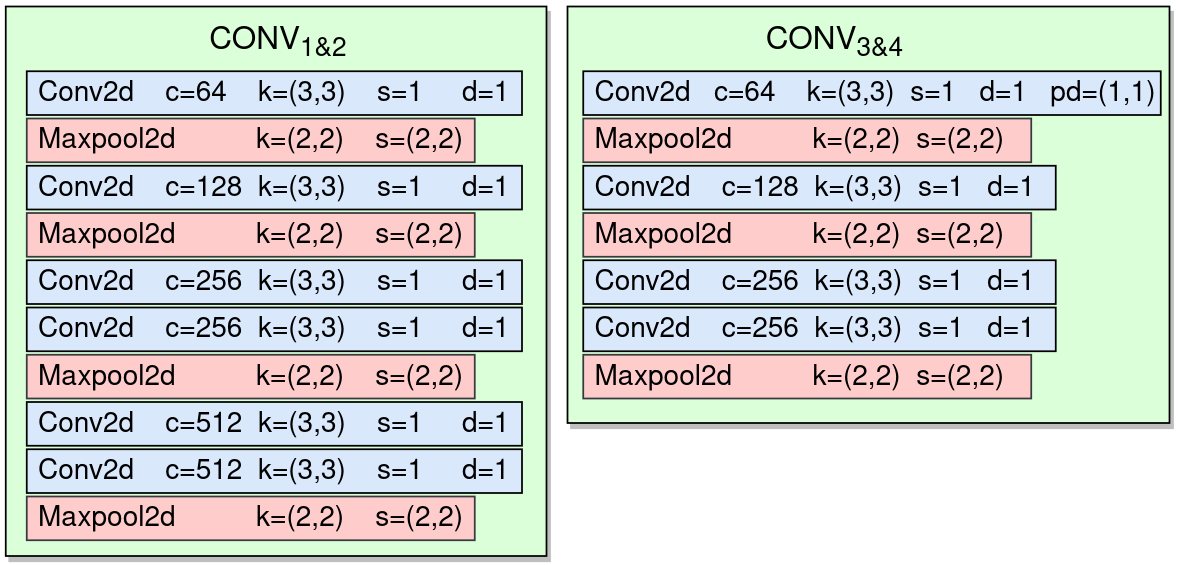}}
 \caption{CNN structure and layer parameters. The layer parameter names are abbreviated as follows: c=number of channels, k=kernel size, d=dilation, s=stride and pd=padding.}
 \label{fig:cnn}
\end{figure}

Obtained CSMs are fed to the similarity measuring section of the model in order to examine the musical similarity of the input tracks. There are two parallel CNNs processing CSMs from different levels of abstraction. Matrices $C_1$ and $C_2$ are fed to CONV\textsubscript{1\&2} while $C_3$ and $C_4$ are processed in CONV\textsubscript{3\&4}. The details about the inner structures of CONV\textsubscript{1\&2} and CONV\textsubscript{3\&4} are provided in Fig. \ref{fig:cnn}. The above-mentioned approach of was chosen rather than using one shared CNN for each $C_k$ as was done in method \cite{refe1}. Given the significant differences in the sizes of $C_1\in\mathbb{R_+}^{194 \times 194}$, $C_2\in\mathbb{R_+}^{93 \times 93}$, $C_3\in\mathbb{R_+}^{43 \times 43}$ and $C_4\in\mathbb{R_+}^{37 \times 37}$ it seems counter-intuitive to treat the CSMs without taking the matrix resolution into account. Also the fact that $C_3$ and $C_4$ tend to contain more diagonal patterns for inputs of class 1 suggests that it is reasonable to train two separate CNNs that can detect level-specific data patterns.

The CNNs are followed by four fully connected neural network layers, each dedicated for processing CSMs of a certain level. The number of neurons at each branch are 32768, 2048, 1024 and 1024 which is equal to the number of pixel values remaining in the convolved CSMs. There are four hidden neurons following the first layer, each connecting the neurons at the same branch to enable different weighing for each $C_k$. The hidden neurons are connected to a single output neuron. Finally, a sigmoid function is applied to the last neuron to produce the similarity score $\hat{Y}$ as the model output. More details about the structure of the CNNs and their parameters are shown in Fig. \ref{fig:cnn}.

\subsection{Training parameters}
The model development was done using PyTorch. Batch training with a batch size of 8 was used in the training. AMSGrad optimizer algorithm was used with an initial learning rate of 0.001 \cite{amsgrad}. The learning rate was decreased gradually during the training by using a learning scheduler with a factor of 0.1 and a patience of 5. The maximum number of training epochs was set to 50. The epoch with the minimum validation loss was considered as the optimal state for the trainable parameters. Some of the values for the hyperparameters were obtained by trial and error. 

\section{Data and alternative feature extraction methods}\label{sec:data}

\textit{Second Hand Songs 100K} dataset (SHS100K) was used in the training of the model \cite{shs100k}. Originally SHS100K contains about 10k songs and 100k audio tracks with a diverse range of music genres, artists and song interpretations. After discarding all the unavailable tracks the number of songs and tracks was reduced to 6.2k and 67.8k, respectively. Pre-processing of the SHS100K data was done as follows. Samples of class 1 were first created by forming pairs of tracks presenting the same underlying song. The maximum number of cover versions about the same song was set to 25. For songs with less then 25 tracks of them, additional track pairs were obtained by linking cover versions of same song with each other. A total amount of 96.5k track pairs was obtained and the data was split into a training set and a holdout validation set with a division of 80\%-20\%. In the data split, all track pairs presenting the same song were put exclusively in the same set. Samples of class 0 were then generated within the two sets by shuffling all the cover versions to pair tracks presenting different songs. During the training the samples of class 0 were recreated after every epoch in order to increase the number of new examples. A class imbalance with a ratio of 1:3 was used between classes 1 and 0, respectively.

Three alternative features extraction methods were applied to the data. The variations are named as 1) basic method, 2) chorus alignment method and 3) crowd noise method. Method-specific features were extracted and used in the training and validation of three separate models to study the effects of the variations on the system performance. The basic method and its feature extraction was done according to the steps explained in Section \ref{feat}. In the chorus alignment method, a pre-trained chorus detection algorithm was used to determine the placement of the examined audio segment within the audio tracks \cite{deepchorus}. The timestamps of the choruses in each song were estimated based on the outputs of the chorus detection algorithm and the CQ-features were then extracted starting from the first identified chorus. Songs without any predicted choruses were dealt as in the basic method. In cases where less than two minutes of audio was available after the predicted start time of a chorus, the last two minutes of the track was used. In the crowd noise method the basic features were mixed with sounds of crowd noise from real-life music events. In total, 54 minutes of crowd noise data was gathered from 25 Youtube videos. The recordings include sounds of applause, ambient stadium noise and crowd cheering with varying crowd sizes and venues. The mixing of the crowd noise recordings and the song tracks was done in frequency domain as an element-wise addition of the extracted CQ-spectrograms. Crowd noise was applied on 50\% of the training and validation data. During the training, the mixing of crowd noise was randomized before each epoch while in validation the noise was only applied once. Varying delays $d = [-15 s, \dots, 117 s]$ and gains $G \in \{-6 $ dB$, -9$ dB$, -12$ dB$\}$ were applied to the crowd noise before the mixing. The presence of crowd noise and the amount of delay and gain were also randomized between the training epochs.


A custom live music dataset was used in the model evaluation. The custom live music dataset was manually collected from a local library's cd-album collections\cite{piki}. The dataset consists of 270 original studio versions of songs and 405 live recording tracks where each live track presents one of the original studio versions. A variety of music genres and eras are included in the dataset with the main focus being on western popular music. In most of the live recording tracks the performing artist is the same as in their corresponding studio version. A full list of the songs and artists in the custom live music dataset is presented in a related thesis \cite{dippa}. Covers80 dataset was also used in the model evaluation as a separate experiment \cite{c80}. Covers80 is a benchmark dataset that consists of 80 songs, each performed by two artists. The results with Covers80 dataset provide a common ground for comparing the model performance against the existing CSI-methods in the literature. Information about the presented datasets is shown in Table \ref{tab:data}.

\begin{table}[htbp]
 \begin{center}
 \begin{tabular}{|c||c|c|p{1,3cm}|}
  \hline
  & SHS100K & Covers80 & Live Music Dataset \\
  \hline\hline
  Audio tracks & 59,6k & 160 & \hfil678 \\
  Song pairs & 55,4k & 80 & \hfil405 \\
  Artists & 31,3k & 92 & \hfil37 \\
  Live music & No* & No & \hfil Yes \\
  Audio format & mp3 & mp3 & \hfil wav \\
  Raw audio & 340 GB & 152 MB & \hfil28 GB \\
  Features & 8 GB & 20 MB & \hfil91 MB \\
  \hline
 \end{tabular}
\end{center}
 \caption{Statistics and information about the used datasets.}
 \label{tab:data}
 \vspace{-4mm}
\end{table}

\section{Evaluation}\label{sec:evaluation}
The performance of the three methods are measured using the common evaluation metrics from the task of CSI \cite{mirex}. The used metrics are \textit{Precision at 10} (P@10), \textit{the mean rank of the first correctly identified cover} (MR1) and  \textit{Mean Average Precision} (MAP). The final results of the experiments are shown in Table \ref{tab:results}. For each query the number of relevant items is 1. Therefore, the best possible values for $P@10$ and $MAP$ are 0.100 and 1.000, respectively.


\begin{table}[h]
 \begin{center}
 \begin{tabular}{|l|lll|}
  \hline
  Custom live music dataset & P@10 & MR1 & MAP \\
  \hline
  Basic & 0.096 & 4.007 & 0.902 \\
  Chorus alignment& 0.094 & 5.627 & 0.868 \\
  Crowd noise& 0.096 & 4.232 & 0.905 \\
  \hline
  \hline
  Covers80 & P@10 & MR1 & MAP \\
  \hline
  Basic     & 0.075 & 9.6 &  0.641 \\
  Chorus alignment& 0.077 & 9.713 & 0.614 \\
  Crowd noise& 0.075 & 9.438 & 0.586 \\
  \hline
 \end{tabular}
\end{center}
 \caption{The results of the experiments evaluated with Custom live music dataset and covers80 dataset}
 \label{tab:results}
 \vspace{-4mm}
\end{table}

The basic method and the crowd noise method perform well on the custom live music dataset given that the training was done mostly without live music data. When further analysing the live music experiments, the basic method succeeds in 87.4\% of the queries by retrieving the desired studio version as the most relevant result. Moreover, 93.6\% of the queries retrieve the corresponding studio version among the five highest-ranked predictions. The crowd noise method provides very similar results and is able to identify 87.4\% of the live music tracks while 92.8\% of the queries retrieve the desired song track among the top-5 predictions. The chorus alignment method has slightly worse performance compared to the others. This is because in some cases the chorus alignment algorithm tends to increase the alignment mismatches rather than synchronizing them as was intended. It is worth noting that the detection accuracy of the chorus alignment algorithm is not perfect to begin with, and that the pre-trained network has not been trained for live music use.


The results of the experiments with Covers80 dataset are relatively poor compared to the results obtained with the custom live music dataset. This can be assumed to be a result of more remarkable musical differences between the samples of call 1 in the two datasets. The experiment proves that more musical changes should be introduced in the features extraction stage. Simulated song key transpositions and tempo alterations are examples of simple and effective ways to introduce more variance in the data \cite{refe1}.  


The effects of alternative loss functions remain unknown. One problem with BCE-loss is that it is blind to relative differences between song similarities. BCE-loss considers all of the examples equally similar or dissimilar which in reality is a false assumption. As a result, while the majority of the live recordings are identified correctly, some of the queries provide a very low similarity score and thus decrease the obtained evaluation metrics. Triplet Loss and Lifted Structure Loss are examples of loss functions that could take this relative separation into account.

\section{Conclusion}\label{sec:conclusion}
This paper studies the problem of live music song identification. A live music song identification system was designed and implemented by applying similarity learning in training of a SCNN-based model that uses CSMs of multi-level deep sequences. In related literature similar methods are used in the task of CSI and in this study the same techniques were applied with the focus on live music data. A custom audio dataset of live music recordings was collected manually for the evaluation of the model. Three different feature extraction methods were examined in order to see the how they affect the model performance. After training the model with cover songs and evaluating with live music data, the best system retrieved the correct studio version for 87.1\% of the given live recording queries. The obtained results are promising considering the nature of the training data.

\bibliographystyle{IEEEtran}
\bibliography{refet}

%
%
%
%
%

\end{document}